\documentclass[structabstract]{aa}  
\usepackage{graphicx}
\usepackage{txfonts}
\usepackage{amssymb}
\usepackage{amsmath}
\usepackage{float}
\usepackage{times}
\usepackage[markup=default]{changes}
\usepackage{soul}
\usepackage[flushleft]{threeparttable}

\usepackage{color}

\def\msun{M$_{\odot}$}

\def\igr{IGR\,1955+0044}

\def\it{\sl}
\def\degs{\ifmmode ^{\circ}\else$^{\circ}$\fi}
\def\amin{\ifmmode ^{\prime}\else$^{\prime}$\fi}
\def\asec{\ifmmode ^{\prime\prime}\else$^{\prime\prime}$\fi}

\def\m{$^{\rm m}$}

\def\degs{\ifmmode ^{\circ}\else$^{\circ}$\fi}
\def\amin{\ifmmode ^{\prime}\else$^{\prime}$\fi}

\def\eqalign#1{\null\,\vcenter{\openup1\jot \m@th
   \ialign{\strut\hfil$\displaystyle{##}$&$\displaystyle{{}##}$\hfil
   \crcr#1\crcr}}\,}
\usepackage{color}

\begin{document}

   \title{IGR J19552+0044: A new asynchronous short period polar}
   \subtitle{Filling the gap between intermediate and ordinary polars}
   \titlerunning{IGR J19552+0044}
   \authorrunning{G. Tovmassian et al.}

   \author{ G.~Tovmassian\inst{1},
        D.~Gonz\'alez-Buitrago\inst{1,2},
        J.~Thorstensen\inst{3},
        E.~Kotze\inst{4},
        H.~Breytenbach\inst{4},
        A.~Schwope\inst{5},
        F.~Bernardini\inst{6},
        S.~V.~Zharikov\inst{1},
        M.~S.~Hernandez\inst{1,7},
        D.~A.~H.~Buckley\inst{4},
        E.~de~Miguel\inst{8,9},
        F.-J.~Hambsch\inst{8},  
        G.~ Myers\inst{8},   
        W.~Goff\inst{8},
        D.~Cejudo\inst{8},
        D.~Starkey\inst{8},
        T.~Campbell\inst{8},
        J.~Ulowetz\inst{8},
             W.~Stein\inst{8},
             P.~Nelson\inst{8}
             D.~E.~Reichart\inst{10}, 
             J.~B.~Haislip\inst{10}, 
             K.~M.~Ivarsen\inst{10},  
             A.~P.~LaCluyze\inst{10}, 
             J.~P.~Moore\inst{10}
   \and          
   A. S. Miroshnichenko\inst{11}   
             }

   \institute{Universidad Nacional Aut\'{o}noma de M\'{e}xico, Instituto de Astronomia,
Unidad Academica en Ensenada, Baja California, C.P. 22860, M\'{e}xico .   
                 \email{gag@astro.unam.mx} 
\and 
              Department of Physics and Astronomy, University of California, Irvine, California 92697, USA
\and
              Department of Physics and Astronomy, Dartmouth College, Hanover NH 03755, USA
\and
              South African Astronomical Observatory, PO Box 9, Observatory 7935, Cape Town South Africa                          
\and       
               Leibniz-Institut f\"ur Astrophysik Potsdam (AIP), An der Sternwarte 16, D-14482 Potsdam, Germany
\and
              New York University Abu Dhabi, P.O. Box 129188, Abu Dhabi, United Arab Emirates    
\and
             Instituto de F\'{i}sica y Astronom\'{i}a, Facultad de Ciencias, Universidad de Valpara\'{i}so, Av. Gran Breta\~{n}a 1111 Valpara\'{i}so, Chile             
\and
                Center for Backyard Astrophysics (CBA)              
\and
                Departamento de Ciencias Integradas, Facultad de Ciencias Experimentales,
Universidad de Huelva, 21071 Huelva, Spain
\and            
                Department of Physics and Astronomy, University of North Carolina at Chapel Hill,  Chapel Hill, NC 27599, USA
\and            
                Department of Physics and Astronomy, University of North Carolina at Greensboro, Greensboro, NC 27402-6170, USA                          
                 }

   \date{Received ; accepted }
 \abstract{Based on {\sl XMM--Newton} X-ray observations  IGR\,J19552+0044 appears to be  either a pre-polar or an asynchronous polar. }
 {We conducted  follow-up optical observations to identify the sources and  periods of variability precisely and to classify this X-ray source correctly. }{Extensive multicolor photometric and medium- to high-resolution spectroscopy observations were performed and period search codes were applied to sort out the complex variability of the object.}{We found firm evidence of discording spectroscopic ($81.29\pm0.01$\,m) and photometric ($83.599\pm0.002$\,m) periods that we ascribe to the white dwarf (WD)\ spin period and binary orbital period, respectively. This confirms that  IGR\,J19552+0044 is an asynchronous polar. Wavelength dependent  variability  and its continuously changing shape point at a cyclotron emission from a magnetic WD with a relatively low magnetic field below 20 MG.}
 {The difference between the WD spin period and the binary orbital period proves that IGR J19552+0044  is a polar with the largest known degree of asynchronism (0.97 or 3\%).}
 

   \maketitle
%

\begin{table*}[t]
\begin{center}
\footnotesize
\caption{\small Log of spectroscopic observations}
\begin{tabular}{lllllll|}
\hline\hline

 Date                      &HJD & Telescope+instrument  &  Filter & t$_{\mathrm{exp}}$ & Total time \\
                              & 2450000+       &                                      &            &                s  &                 ks  \\\hline
2011/09/17-18-19   && SPM 0.84m+MEXMAN   &  V    & 90/60             & 25.74 \\
2011/09/26-27-28   && SPM 0.84m+MEXMAN   &  UBVRI    & 60/30/20             & 24.33 \\
2012/07/23-24-26   && SPM 0.84m+MEXMAN   &  I    & 120             & 30.36  \\
2013/07/07-08-13   &6480.841-86.968& SPM 1.50m+RATIR   &  r-i     & 60             & 20.50  \\
2013/07/14-15-16   &6487.914-89.965& SPM 1.50m+RATIR   &  r-i     & 60             & 20.29  \\
2013/07/17-18-24   &6490.838-97.923& SPM 1.50m+RATIR   &  r-i     & 60             & 28.35  \\
2013/07/25-27-28   &6498.820-01.967& SPM 1.50m+RATIR   &  r-i     & 60             & 34.38  \\
2013/07/29-30-31   &6502.791-04.950& SPM 1.50m+RATIR   &  r-i     & 60             & 29.86  \\
2013/08/01-02-03   &6505.843-07.950& SPM 1.50m+RATIR   &  r-i     & 60             & 32.61  \\
2013/08/04-05-06   &6508.841-10.933& SPM 1.50m+RATIR  &  r-i     & 60             & 37.16  \\
2013/08/12-13-14   &6516.782-18.916& SPM 1.50m+RATIR   &  r-i     & 60             & 35.77  \\
2013/08/15-16-17   &6519.770-21.913& SPM 1.50m+RATIR   &  r-i     & 60             & 36.66  \\
2013/06/19-21-24   &6462.604-62.642& PROMPT Apogee Alta   &  V-I     & 120/120             & 70.79  \\
2013/06/26-29-30   &6469.675-73.931& PROMPT Apogee Alta   &  V-I     & 120/120             & 73.91  \\
2013/07/01-02-03   &6474.708-76.931& PROMPT Apogee Alta   &  V-I     & 120/120             & 140.44  \\
2013/07/04-07-08   &6477.787-81.843& PROMPT Apogee Alta   &  V-I     & 120/120             & 61.50  \\
2013/07/09-12-13   &6482.743-86.637& PROMPT Apogee Alta   &  V-I     & 120/120             & 48.00  \\
2013/07/14-16-17   &6487.821-90.678& PROMPT Apogee Alta   &  V-I     & 120/120             & 65.34  \\
2013/07/18-19-21   &6491.525-94.666& PROMPT Apogee Alta   &  V-I     & 120/120             & 81.46  \\
2013/07/23-24-25   &6496.517-98.875& PROMPT Apogee Alta   &  V-I     & 120/120             & 50.54  \\
2013/07/23-24-25   &6496.517-98.875& PROMPT Apogee Alta   &  V-I     & 120/120             & 85.43  \\
2013/07/28-29   &6501.499-02.652& PROMPT Apogee Alta   &  V-I     & 120/120             & 34.88  \\
2013/08/04-06-07   &6508.672-11.580& PROMPT Apogee Alta   &  V-I     & 120/120             & 54.71  \\
2013/08/12-13-14   &6516.782-18.892& PROMPT Apogee Alta   &  I     & 120/120             & 31.05  \\
2013/08/15-16-17   &6519.770-21.877& PROMPT Apogee Alta   &  I     & 120/120             & 24.59  \\
2014/07/02-07-09   &6841.481-47.59& SAO 1.9m + SHOC   &  without    & 0.3   & 25.84  \\ 
\hline
\end{tabular}
\label{tab1}
\end{center}
\begin{tabular}{l}
\end{tabular}
\end{table*}

\begin{table*}[t]
\begin{center}
\footnotesize
\caption{\small  Log of photometric observations}
\begin{tabular}{lllllll|}
\hline\hline

 Date                             &HJD                                  & Telescope+instrument  &  $\lambda$  & t$_{\mathrm{exp}}$ & Total time \\
                                     &2450000+                          &                                          &   range (\AA) &  s    &  ks  \\\hline
2011/06/28-29-30      &5740.856-42.900&   SPM 2.1m+B\&Ch                & 4140-5360       &  600                       & 19.8 \\
2011/07/01               &5743.788-43.850 &   SPM 2.1m+B\&Ch            & 4140-5360       &  600                       & 5.40 \\
2011/09/18-19-20          &5822.646-24.772&   SPM 2.1m+B\&Ch            & 4120-5340       &  600                       & 27.0 \\
2011/09/21-27             &5825.638-31.751&   SPM 2.1m+B\&Ch            & 3490-8100       &  600                       & 16.4 \\
2012/07/17-18-21-25      &6125.695-33.749&   SPM 2.1m+B\&Ch              & 4000-8000     &  600/900               & 34.0 \\
2013/06/28-29-30      &6471.801-73.964&   SPM 2.1m+B\&Ch                 & 4000-8000     &  600               & 28.0 \\
2013/10/31               &6596.597-96.621&   MDM 1.3+ModSpec/Ech         & 4300-7500     &  480                       & 2.40 \\
2013/09/11-12-13         &6546.698-48.617&   MDM 2.4+ModSpec/Tmp         & 4660-6700     &  480                       & 4.80 \\
2013/09/14-15-17         &6549.613-52.612&   MDM 2.4+ModSpec/Tmp         & 4660-6700     &  360                       & 7.80 \\
2013/06/07-08-09         &6450.968-52.801&   MDM 2.4+ModSpec/Tmp         & 4660-6700     &  480/360              & 6.36 \\
2013/06/10-11            &6453.964-54.967&   MDM 2.4+ModSpec/Tmp         & 4660-6680     &  480/360              & 1.80 \\
2013/03/01-02-04         &6352.997-56.034&   MDM 2.4+ModSpec/Tmp         & 4660-6550     &  600/480              & 7.56 \\
2012/12/13               &6274.584-74.599&  MDM 1.3+ModSpec/Tmp  & 4670-6680         &  600                       &  3.60 \\\hline
\end{tabular}
\label{tab2}
\end{center}
\begin{tabular}{l}
\end{tabular}
\end{table*}

\section{Introduction}

AM Herculis stars, or polars,  are  close interacting binaries possessing white dwarfs (WD) with the strongest superficial magnetic fields among cataclysmic variables (CVs) \citep{1995ASPC...85....3W}. The intensity of this field varies from $\sim$10 to 200 MG and is enough to prevent the formation of an accretion disk  and channel the incoming matter  from a late-type companion through the magnetic lines to the magnetic pole(s) of the WD. The WD intense magnetic field and its extended magnetosphere are thought to interact with the magnetic field of the late-type companion star and synchronize the spin period of WD 
with the orbital period of the binary, thereby  overcoming the spin-up torque exerted by the accreting matter 
\citep{1985MNRAS.215..509C,1991MNRAS.250..152K}.
A subset of CVs known as intermediate polars (IPs) or DQ Herculis stars contain WDs  possessing less intense magnetic moments and these CVs do not achieve synchronization  \citep{2004ApJ...614..349N}. 

Among the subclass of polars, there are seven slightly asynchronous systems with $P_{\mathrm {spin}} / P_{\mathrm {orb}}=1-2\%$. These are 
V1432 Aql (RXJ 1940-10), BY Cam, V1500 Cyg, CD Ind (RXJ 2115-58), and Paloma (RX J0524+42) \citep{1999A&A...343..132C,2004ASPC..315..230S,2007A&A...473..511S}. Another asynchronous polar (AP) was discovered by \citet{2016arXiv161104194R} while we were preparing this
paper.  V1432 Aql is the only AP that has a spin period longer than the orbital period, while others have have $P_{\mathrm {orb}} - P_{\mathrm {spin}} \le $ 0.018 P$_{\mathrm {orb}}$ \citep{2004ApJ...614..349N,2016MNRAS.458.1833P}. The exact reason of  the asynchronism is not known yet. Nova eruptions are considered one of the possible culprits \footnote{Among a few APs at least one (V1500 Cyg) definitely has undergone a nova explosion; this AP is also known as  Nova Cygni 1975} \citep{1999A&A...343..132C}, but the efforts to find nova shells around other APs have not been successful so far \citep{2016MNRAS.458.1833P}. It is also assumed that these systems  gain synchronization relatively quickly as shown in the case of V1432 Aql \citep{2014SASS...33..163B}. Very recently, \cite{2016MNRAS.tmp..737H} reported that V1500 Cyg,  which has been known to have large  2\% disparity of its orbital (photometric) and spin (circular polarization)  periods \citep{1988ApJ...332..282S},  has already achieved  synchronization.

IGR J19552+0044 (\igr\ hereinafter) was identified as  a  magnetic CV by \cite{2010A&A...519A..96M} based on follow-up optical spectroscopy of hard X-ray sources detected by {\it INTEGRAL} \citep{2006ApJ...636..765B}. 
\citet{thorhalpern13} obtained time series of spectroscopic and photometric data, but the coverage was insufficient to determine the period in either domain without ambiguity. 
\cite{2013MNRAS.435.2822B} studied the X-ray behavior of the object using {\sl XMM-Newton}.  These authors point out that \igr\ is a highly variable X-ray source with a rather hard spectrum, showing also near-infrared and infrared variability. They inferred a high $0.77$\,\msun\  mass for the WD and a low accretion rate. Their period analysis was inconclusive as to whether the detected periods were orbital or spin. Based on detection of  hard X-ray spectrum and multiple periodicities they proposed the AP nature for the object.   
We conducted follow-up spectroscopic and photometric optical  observations of  \igr. 
We incorporated  \citet{thorhalpern13} spectral observations into our study to expand the time baseline. Here we report the results of this study,  deducting the binary basic parameters (e.g., spin period, orbital period, and magnetic field intensity).

Details of the observations are provided in Section\,\ref{observations}. We present an analysis of the optical spectroscopy and photometry in Section\,\ref{results}. 
We discuss the nature of the system in Section\,\ref{discus}, and conclusions are summarized in Section\,\ref{conclusions}.

\begin{figure}[t]
\setlength{\unitlength}{1mm}
\resizebox{11cm}{!}{
\begin{picture}(100,65)(0,0)
\put   (2,0)         {\includegraphics[width=7cm,bb=0 150 580 720,clip]{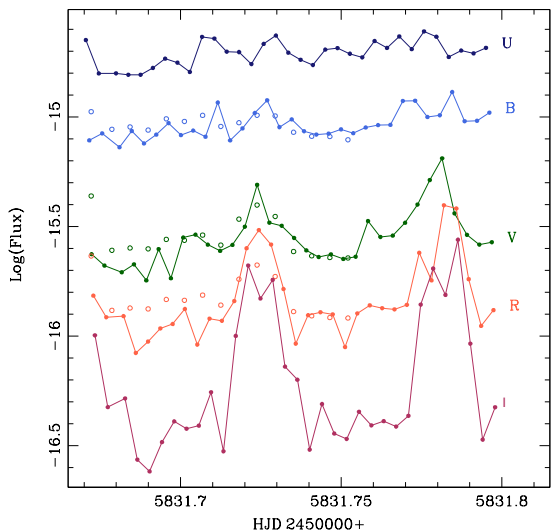}}
\end{picture}}
      \caption{UBVRI photometry of \igr. A time series (quasi-simultaneous)  in all five filters during a night in September 2011 are plotted with connected filled circles. The open points were inferred from the spectrophotometry obtained at the same time by integrating flux in the portions of spectra corresponding roughly to  $BVR$ filters. 
 Spectrophotometric measurements taken with a long slit 
 demonstrate satisfactory flux calibration. The flux is given in ergs cm$^{-2}$ s$^{-1}$ $\AA^{-1}$.
         }
         \label{fig01}
   \end{figure}


\section{Observation and data reduction}
\label{observations}

The time-resolved CCD photometry and long-slit spectral observations of \igr\  were obtained on the 0.84 m, 1.5 m and 2.1 m telescopes of the Observatorio Astron\'{o}mico 
Nacional at San Pedro M\'{a}rtir (SPM) in Mexico. On September 26, 2011 we performed simultaneous spectroscopic and UBVRI photometric observations using the 2.1\,m telescope with B\&Ch spectrograph and  the 0.84 m with the MEXMAN filter wheel. We observed the source for three years using different combination of telescopes and instruments. The bulk of data   were obtained in Bessel  V, I  and SDSS r, i -- bands   using the 0.84m/MEXMAN  and the 1.5 m/RATIR telescope/instrument, respectively.   Landolt photometric stars  were also observed for the absolute calibration. Exposure times were 60\,s for the RATIR observations and ranged from 20\,s to 90\,s,  depending on the filter and conditions for the 0.84\,m telescope. The images were bias-corrected and flat-fielded before the differential aperture photometry was carried out.  The errors of the  CCD photometry were calculated from the dispersion of the magnitude of the comparison stars. 

We launched the monitoring of \igr\  using two 0.4 m robotic PROMPT telescopes located in Chile \citep{2005NCimC..28..767R}.
In a two month campaign from June 19 to August 7, 2013   the PROMPT telescopes were intensely employed.
Most of the  observations in 2013 were performed in the I filter, whereas at the beginning of June we also gathered some observations in V filter. 
The exposure times were 120 s throughout the campaign. 

A portion of the photometric data included in this paper were obtained by observers
of the Center for Backyard Astrophysics (CBA), which is a global network of telescopes 
devoted to the observation of cataclysmic variables \citep{1993ApJ...417..298S,2016MNRAS.457.1447D}.
Typical apertures are in the 0.25-0.40 m range.
A total of 10 CBA observatories contributed to this campaign, providing 
~480 hours of time-series photometry. Most of the data was unfiltered, with
exposure times ranging from 45 to 60 sec.

Additionally, we obtained high time resolution photometry (0.25--3\,s integrations) of \igr\  without filter (in white light). We used the 1.9\,m telescope of South African Astronomical Observatory (SAAO) equipped with  the SHOC camera. These observations were carried out as part of a program to search  and study  high-frequency quasi-periodic oscillations.  
generated in the accretion columns in CVs.

 \begin{figure}[t]
 \setlength{\unitlength}{1mm}
\resizebox{11cm}{!}{
\begin{picture}(100,65)(0,0)
\put   (-5,0)    {\includegraphics[width=3.6in, bb=100 230 540 540,clip]{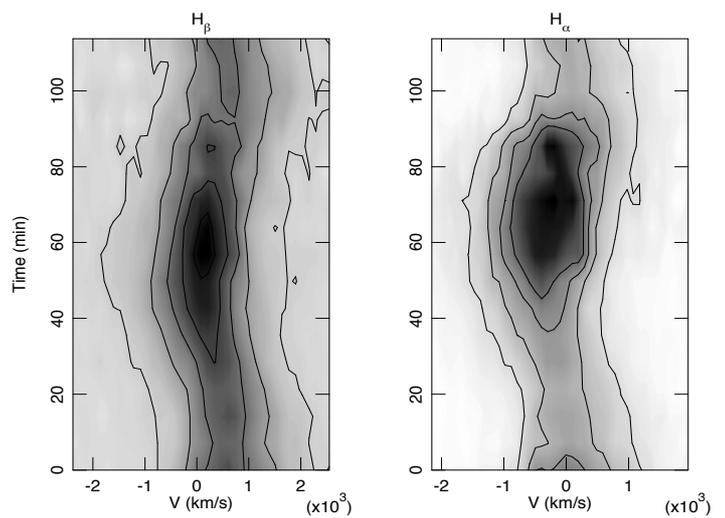}}
\end{picture}}

      \caption{  
      Trailed spectra of \igr\ centered at the H$_\alpha$\ and H$_\beta$\  are shown in the right and left panels, respectively. The bi-dimensional image is composed using 19 continuous, evenly spaced spectra obtained during about two hours on September 15, 2013. Two components are clearly visible in the right panel. The narrow component is stronger at some orbital phases.      }
         \label{fig11}
   \end{figure}

The spectroscopic observations were conducted with the Boller and Chivens spectrograph, equipped with a 13.5 $\mu$m ($2174 \times 2048$) Marconi E2V-4240 CCD chip, using the 2.1 m telescope. A portion of the observations were obtained  with a 1200\,l/mm grating to study profiles of emission lines (FWHM$=2.1\AA$); some observations were made using a 300\,l/mm (FWHM$=8\AA$) grating to cover almost the entire optical range and study cyclotron lines. The   wavelength calibration was made with an Cu-Ne-Ar arc lamp. The spectra of the object were flux calibrated using spectrophotometric standard stars observed during the same night. The low-resolution spectra were obtained with a wide slit $350 \mu m$ to improve the flux calibration of spectra taken without slit orientation along the parallactic angle. 

We also included spectra with both the 2.4 m Hiltner and 1.3 m McGraw-Hill telescopes at MDM Observatory on Kitt Peak,
Arizona.  We used the modspec spectrograph with either
the {\it Echelle} or {\it Templeton} CCD detectors.  These
SITe chips are identical in pixel size and therefore both yield 2.0
\AA\ pixel$^{-1}$, but the Echelle has a larger format and
covers a greater spectral range.  \citet{thorhalpern13}
give more detail on the observing and analysis procedures.

Reduction and preliminary analysis of all spectroscopic and the photometric observations from SPM were carried out using long-slit spectroscopic and aperture photometry  packages available in IRAF\footnote{IRAF is distributed by the National Optical Astronomy Observatories,  which are operated by the Association of Universities for Research in Astronomy, Inc., under cooperative agreement with the National  Science Foundation.}.
The logs of  observations are presented in Tables \ref{tab1} and \ref{tab2}.

  \begin{figure}
    \setlength{\unitlength}{1mm}
\resizebox{11cm}{!}{
\begin{picture}(100,75)(0,0)
\put   (2,0) {\includegraphics[width=3.6in, bb=0 150 650 690,clip]{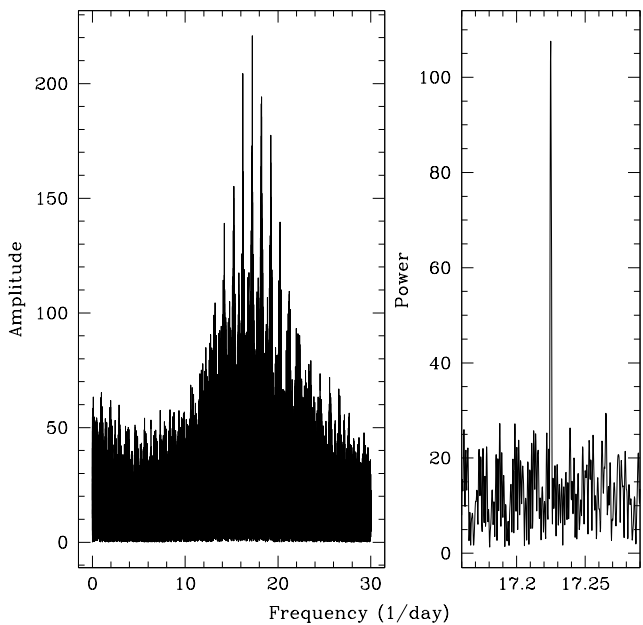}}
\end{picture}}

      \caption{Results of DFT period analysis of RV of emission lines. In the left panel the amplitude vs. frequency of H$_\beta$\ line is presented.  In the right panel the power of the same time series is presented after cleaning for alias frequencies.  The maximum peak is at  frequency $f_{\mathrm {o}}=17.22497$ cycles/day.}
         \label{fig02}
   \end{figure}

\section{Results}
\label{results}

\subsection{Spectroscopic period and spectrophotometric characteristics}

Spectra of \igr\ were previously published by  \citet[see Figure\,6 of][]{2010A&A...519A..96M} and \citet[Figure\,14 of][]{thorhalpern13}. We found that the optical spectra of  \igr\   are consistent with that of a CV, but its particular classification is not simple. The object shows a standard set of hydrogen and helium lines, which are single peaked but with variable profile and intensity. A $15-25$\AA\ FWHM of emission lines present  in the $\lambda\,3800-8050$\,\AA\  range  is typical for CVs. The He\,{\sc ii} line  is prominent, indicating presence of a high-ionization source in the system, but its intensity is less than 1/2 H$_\beta$. In high accretion rate polars the intensity of He\,{\sc ii} 4686 and  H$_\beta$ are often of the same order. No spectral features of the secondary star are visible in the optical spectra.  
The source is highly variable and intensity changes are notable by eye not only in the lines but also in the continuum and, more intriguingly, in the shape of the continuum. 
This variability 
is best demonstrated by the multicolor light curve presented  in Figure\,\ref{fig01}.   
It is also a better visualization of  the scale  and  the wavelength dependency  of the variability.  We discuss the photometric behavior of the system in  Section\,\ref{sec:phot}.

   \begin{figure}[t]
 \setlength{\unitlength}{1mm}
\resizebox{11cm}{!}{
\begin{picture}(100,75)(0,0)
\put   (2,0) {\includegraphics[width=3.4in,  bb=0 140 620 720,clip]{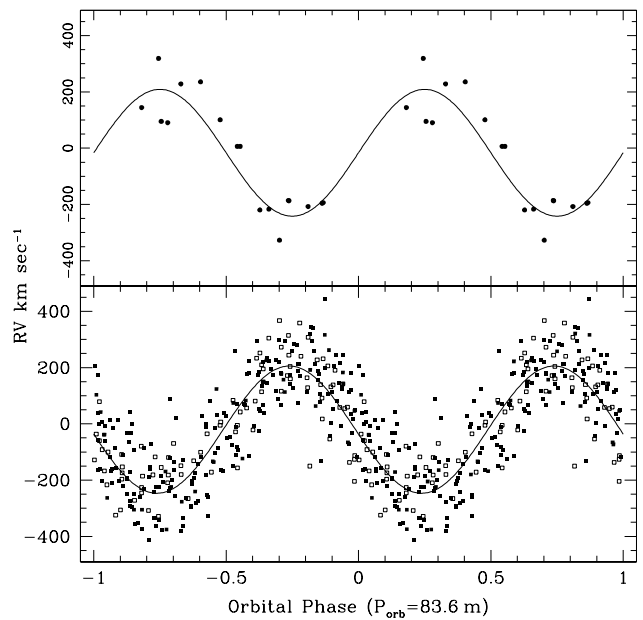}}   
 \end{picture}}         
             \caption{Radial velocity curves of \igr. Bottom panel: Single Gaussian measurements of  H$_\beta$\  fitted with a sine curve are shown. The filled squares are data obtained at SPM; the open squares are from MDM. Top panel: The measurement of the narrow component of the H$_\alpha$\ line wherever we were able to distinguish it in the line profile, is shown.
             The sine curve with fixed period determined from the H$_\beta$\ analysis was used to fit to these points. There is a small phase shift of the  H$_\beta$\ RV curve from being totally opposite to that of H$_\alpha$. }
        \label{fig03}
   \end{figure}

\begin{figure*}[th]
 \setlength{\unitlength}{1mm}
\resizebox{11cm}{!}{
\begin{picture}(100,80)(0,0)
\put   (2,0) {\includegraphics[width=3.2in, bb=0 0 760 720,clip]{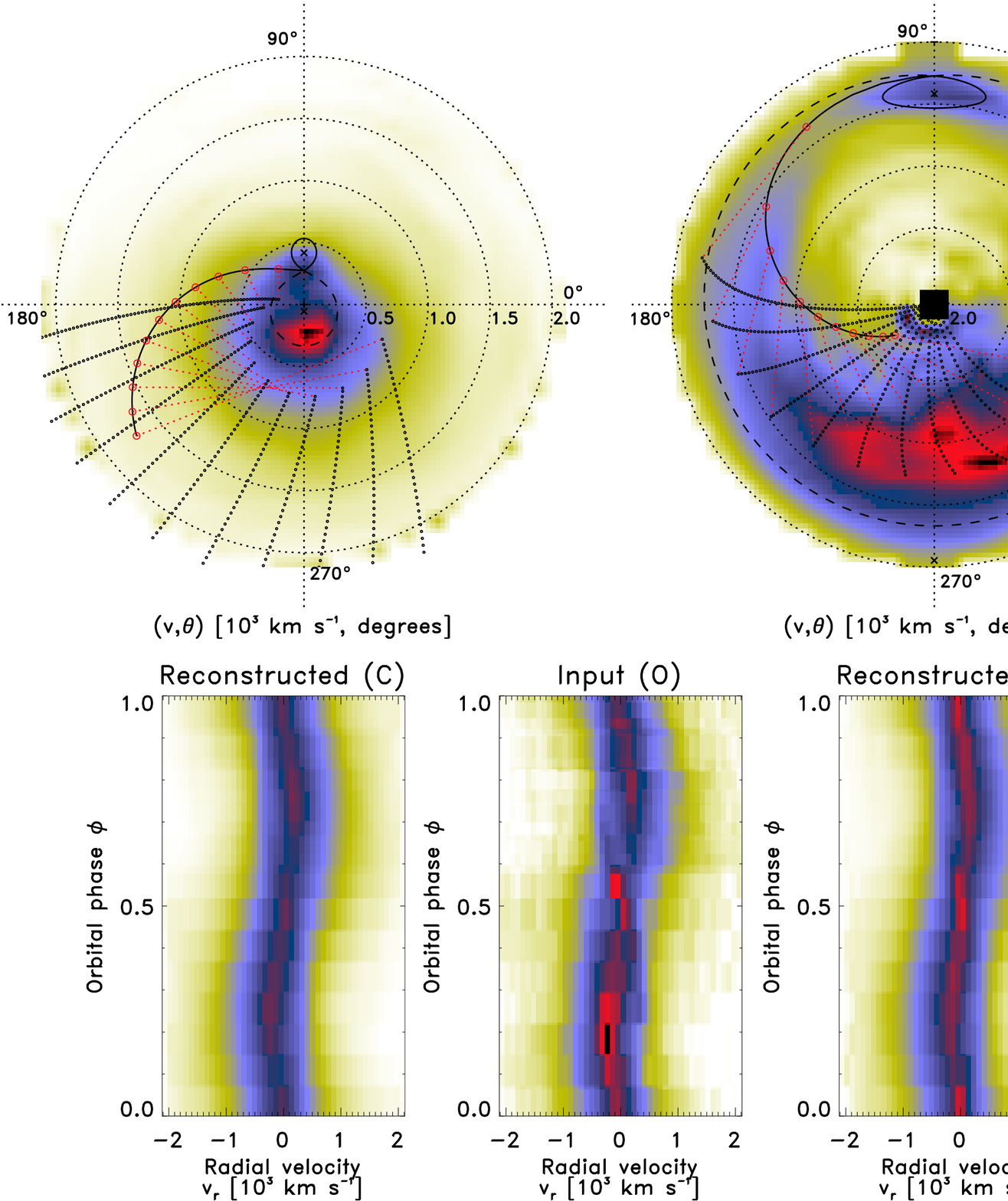}}
\put  (2,4) {\it a)}
\put   (90,0) {\includegraphics[width=3.2in, bb=0 0 760 720,clip]{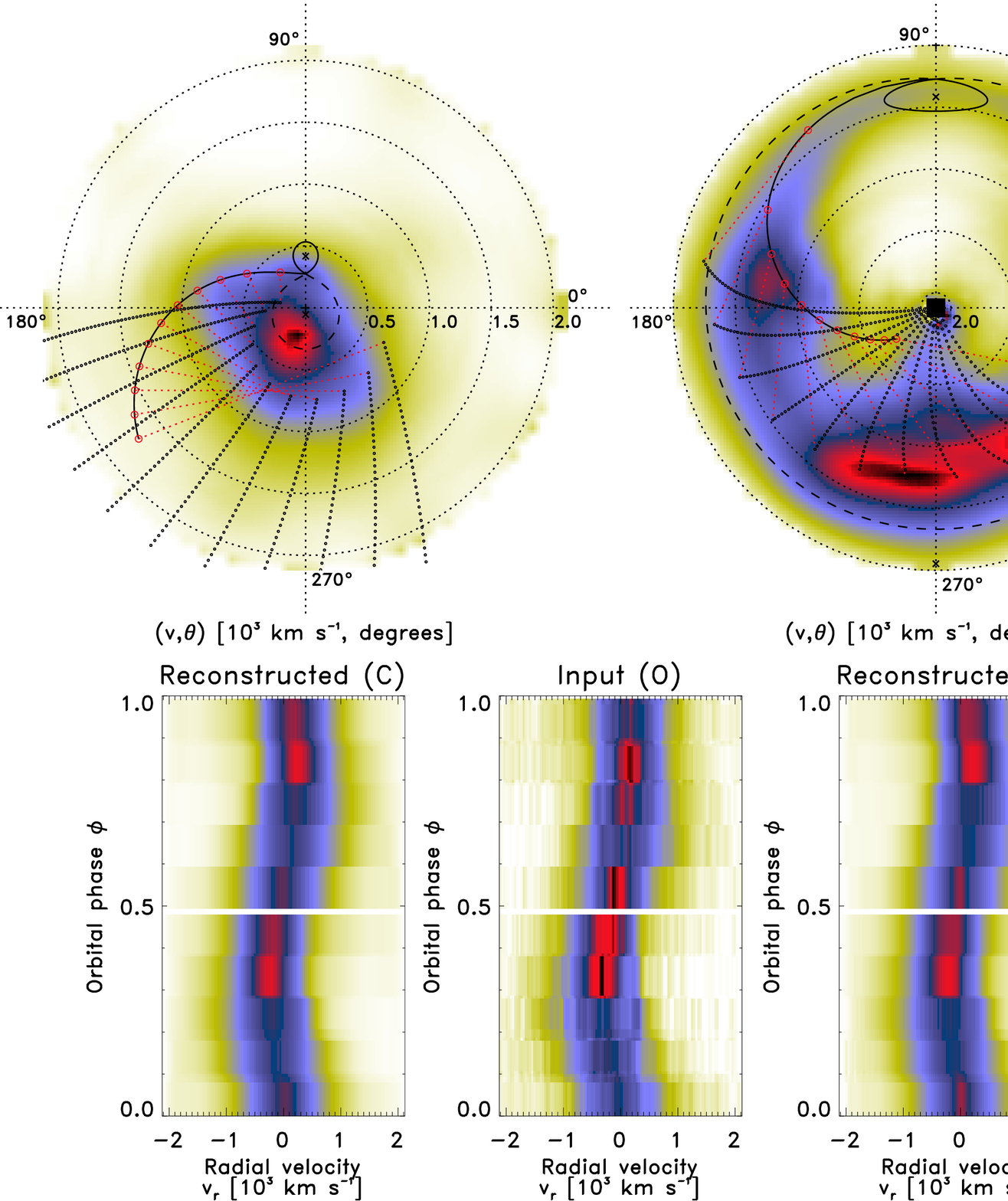}}
\put  (90,4) {\it b)}
  \end{picture}}   
      \caption{Standard and inside-out Doppler tomography based on the H$_\alpha$ emission (a) and the H$_\beta$ emission (b). For each spectral line a pair of maps in the top  panels show the standard tomogram (left side) and the inside-out tomogram (right side). In the bottom panel, three frames accompany maps of each line. They are comprised of the middle frame showing the observed trailed spectra with the reconstructed spectra for the standard and inside out cases to the left and right, respectively.
      See text for details.  
      }
         \label{fig:05}
   \end{figure*}

The trailed spectrum of the H$_\alpha$ line from a single night (September 15, 2013) covering the entire binary period is presented as a  two-dimensional image in  Figure\,\ref{fig11}.  The profiles of emission lines are complex, however  it is not possible to disentangle  these profiles in most  cases. Only H$_\alpha$ clearly shows  a narrow component that can be separated  from the otherwise broader line at some phases.    We attempted to deblend the line  with two Gaussians using the corresponding {\it splot} function in IRAF, but we could resolve two separate components 
in just about  half of the orbital phases.  

Hence, we used the H$_\beta$\ line, as a whole, to determine the spectroscopic period based on the radial velocity (RV) variations.  We chose H$_\beta$\ because it is present in all observed spectra and is the most intense line. The spectral observations span more than 700 days and are comprised of  blocks of several nights each with more than 300  RV measurements in total. The contribution from the narrow component  in H$_\beta$\ is much smaller than in H$_\alpha$, which makes it suitable for the task. Therefore the RVs were measured by fitting a simple Gaussian to the line profile and using its central wavelength. A period search in the RV time series using a discrete Fourier transform (DFT) algorithm implemented in Period04 \citep{2005CoAst.146...53L} reveals a dominant peak at 17.225 cycles/day  corresponding to a 83.6\,min period. The power spectrum is contaminated by aliases generated by  an uneven time series and  beat periods  generated by multiple periodicities. The convolution of the data with the spectral window, also known as a {\sl Clean} procedure \citep{1987AJ.....93..968R} reveals a single peak at the dominant frequency. The existence of this peak is also confirmed by a detection of beat periods in photometry, which we describe later on.  The power  and corresponding {\sl Clean} spectrum are presented in Figure\,\ref{fig02}. 

The RV curves  folded with the determined period are presented in  Figure\,\ref{fig03}.  The bottom panel represents the measurements of H$_\beta$\ and the corresponding sinusoidal fit. There is  a very wide spread of points around the best-fit sine curve. This is partially a result of a poor fit of a Gaussian to the line profile, but is also a consequence of the intrinsic velocity dispersion of the  emitting gas. Usually in CVs the spectroscopic period reflects the orbital motion, but not necessarily of the stellar components, i.e the phase zero does not necessarily correspond to the binary conjunction. Assuming the tentative magnetic CV classification of \igr,\ we may find a large velocity amplitude; it would not be surprising if it exceeded 400 km\,s$^{-1}$. 
The  lines formed in the  mass transfer stream of polars often show much higher velocity amplitudes. The other available lines  were H$_\delta$ and  He\,{\sc ii} 4686, which roughly follow the same pattern as  H$_\beta$.

The object is not eclipsing, hence the orbital conjunction of stellar components is not known at this point.  Measurement of RVs of  the narrow component  help to fetch the zero point corresponding to the inferior conjunction  of the red dwarf component, assuming  that \igr\ is a magnetic CV.  In such a case the narrow component  originates from the irradiated face of the secondary star due to heating by the  X-ray beam from the magnetic pole of the primary \citep{1999MNRAS.304..145H,2016A&A...595A..47K}.  
The RVs of the narrow component of the H$_\alpha$\ line are presented in the top panel of Figure\,\ref{fig03} with the sine fit. According to calculations the +/- crossing of the rest velocity of the system corresponds to  the HJD=2455740.7035  ephemeris.  The narrow and wide (bottom panel) components are nearly, but not exactly, in a counter phase. 

The mass transfer and accretion in polars including APs takes place under strong influence of the magnetosphere of the WD and is very different from the remaining 
CVs \citep[see reviews by ][]{1990SSRv...54..195C,1999MNRAS.310..189F}. 
Observationally,  three distinct components of emission lines were identified in polars \citep{1997A&A...319..894S}.  Not all three components are observable in every polar,  which depends on  the orientation of the magnetic pole, its intensity, and probably some other factors (curtaining, etc.). It is safe to say that in \igr\ the bulk of emission is concentrated near the WD and hence, reflects the fact that the magnetically confined part of the accretion flow is the dominant source of emission lines.  Careful examination of all available spectra reveals that the narrow component is visible not only in H$_\alpha$\, , but  some weak contribution can also be traced to the ballistic part of the accretion stream. This is particularly notable in the H$_\beta$\ line.  

Doppler tomograms, especially their inside-out projections \citep{2016A&A...595A..47K}, help reveal these details. Doppler tomography  in cataclysmic variables was introduced by \citet{1988MNRAS.235..269M}. Traditionally, filtered back-projection inversion or maximum entropy inversion were applied to translate binary-star line profiles taken at a series of orbital phases into a distribution of emission over the binary. Both these methods were primarily designed for interpretation of accretion disk CVs, where the matter is basically confined to the orbital plane. Although the maximum entropy method has also been used successfully for magnetic CVs \citep{2016ASSL..439..195M}, part of the streams and curtains in mCVs are  not in the orbital plane. Hence their interpretation in standard Doppler maps is complicated. 
Recently, \citet{2016A&A...595A..47K}  came up with the so-called inside-out projection to address magnetic CVs specifically. Here we use both, the standard and  inside-out Doppler maps to demonstrate the geometry of the binary system in the velocity space.

Figure \ref{fig:05}  shows the standard and the inside-out Doppler tomography based on the H$_\alpha$\ and H$_\beta$\ emission lines, respectively. The basic structure of the emission components in the observed H$_\alpha$\ and H$_\beta$\ spectra is reproduced in the reconstructed spectra from both the standard and inside-out projections. To aid the interpretation of the emission distribution in the tomograms, we overlay a model velocity profile based on arbitrary but reasonable parameters for a magnetic CV with a $\sim 84$\,m orbital period. The primary mass was set to M$_{\mathrm wd}=0.75$\msun\ \citep{2015SSRv..191..111F}, the mass ratio was set to $q=0.13$\ \citep{2006MNRAS.373..484K}, and the inclination angle was set to $i=65^{\circ}$\ given that the observed RV indicate a high inclination angle, yet no eclipses has been observed. The model velocity profile includes the Roche lobes of the WD (dashed line), the secondary (solid line), as well as a single particle ballistic trajectory from the $L_{1}$ point up to $105^{\circ}$ in azimuth around the WD (solid line). Magnetic dipole trajectories are calculated at $10^{\circ}$ intervals from $5^{\circ}$ to $105^{\circ}$ in azimuth around the WD (thin dotted lines). The dipolar axis azimuth and co-latitude are are taken to be $36^{\circ}$ and $20^{\circ}$, respectively.
In the H$_\alpha$\ tomograms the emission associated with the irradiated face of the secondary is clearly visible in the position traced by the velocity profile of the secondary. This emission, however, is not isolated so clearly in the H$_\beta$\ tomograms. On the other hand, in the H$_\beta$\ tomograms the emission associated with the ballistic stream is more prominent and well traced by the model single particle trajectory.  The most prominent feature in all the tomograms is the brighter emission in their lower halves. From the model velocity profile we deduce that this emission may be associated with the threading region, that is to say, where the matter in the ballistic stream is picked up by the magnetosphere of the WD and is elevated above the orbital plane to channel onto the magnetic pole. This matter 
creates a curtain that appears to be ionized by the energetic beam of the magnetic WD. The emission from the curtain follows the model dipole trajectories toward higher velocities as it is funneled toward the magnetic pole of the WD.  Effectively, the H$_\alpha$\ and H$_\beta$\ tomograms confirm the magnetic nature of the object.
    
\begin{figure}[t]
 \setlength{\unitlength}{1mm}
\resizebox{11cm}{!}{
\begin{picture}(100,75)(0,0)
\put   (2,0)  {\includegraphics[width=3.3in,bb=0 140 620 720,clip]{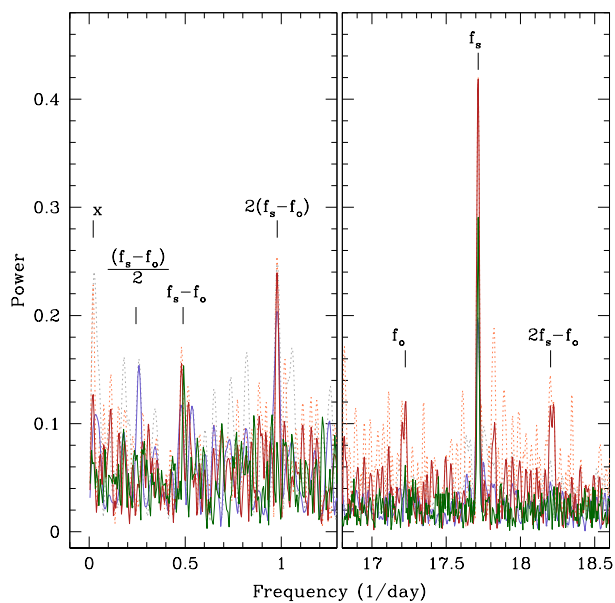}}
  \end{picture}}   
      \caption{Power spectra  multiband photometry of \igr. The solid lines are powers after {\sl clean} procedure is applied to powers presented as dotted lines.  The bluish curves are for the V band, the red are for the I band, and the green are for the WL light curves. }
         \label{fig06}
   \end{figure}

The shape of spectra show spectacular transformation throughout an observing run lasting several hours. However the period of the continuum variability apparently does not coincide with the spectroscopic period determined from the RV variability. This becomes obvious after just two or three individual observational runs. Therefore, we must assess  the photometric periods before returning to this discussion. 

\begin{table}[h]
\begin{center}
\footnotesize
\caption{\small Periods detected in IGR\,J1955+0042}
\begin{tabular}{lllll}
\hline\hline
Period      & Frequency  &  Interpretation & Mode & Amount \\
     min            &       c/d    &                                    &      & points  \\ \hline
 83.60 &   17.22478(40) & orbital &  H$_\alpha$ RV cmp & 303  \\
 83.5996 &   17.22497(40) & orbital &  H$_\beta$ RV  & 303  \\
 83.65 &   17.224 (4) & orbital &  H$_\gamma$ RV  & 107  \\
 83.67     &  17.2(2)       &   orbital     &    WL fast       &  54365 \\  
 81.292  &   17.714(6) &  spin     & I-band phot  & 4658 \\
 81.296    & 17.713(5)    &  spin          & V-band phot  & 2578 \\
 81.288   & 17.7146(3)   & spin          & WL phot   & 32074 \\ 
 81.3       &  17.7(2)       &   spin        &    WL fast       & 54365  \\ 
 1475.4    &   0.976(4) &  2(spin-orb) & I-band phot &  4648 \\
 1473.9    &  0.977(3)     &  2(spin-orb) &  V-band phot &  2578 \\
  3000      &  0.480(4)  &  (spin-orb)   & I-band phot &  4658 \\
 2926.8   & 0.492(6)       & (spin-orb) &  WL phot  & 32074 \\
 82.7       &   17.42(28) &                 &    X-ray$^{*} $     &   \\  
101.6     &    14. 2(3)   &                 &    X-ray$^{*} $      &   \\ 
 \hline
\end{tabular}
\label{tab:fr}
\end{center}
\begin{tablenotes}
\small
\item $^{*} $from \citet{2013MNRAS.435.2822B}
\end{tablenotes}
\end{table}

\subsection{Photometric variability and periods}
\label{sec:phot}

The time-resolved  photometry  confirms  strong variability of the object on different timescales.
The simultaneous multiband photometry shows that the amplitude of the variability depends on wavelength (Figure\,\ref{fig01}). 
We collected sufficient photometric data from various sources to analyze  the complexity of light curves of \igr.  Because of the difference of the amplitude of variability in different photometric bands and the underlying difference of the source of variability, the results in V, I, and white light (WL) are slightly different. 
The power spectra calculated by  Period04 \citep{2005CoAst.146...53L} for light curves in three observed bands (V, I, and WL) are presented in Figure\,\ref{fig06} by short dashed lines. The power spectra after {\sl Clean}-ing to eliminate aliases created by time series are presented by solid lines of similar color.   From the spectrophotometry and simultaneous multicolor photometry, we know that the largest amplitude of variability happens in the I band. Unsurprisingly, the strongest and sharpest peak is detected from I band (plotted in Figure\,\ref{fig06} by red color) at a 17.7 cycles per day frequency corresponding to a 81.3 m period. This matches, within statistical uncertainty, the shortest X-ray periodic signal found by \citet{2013MNRAS.435.2822B}.

The periodic signal is formed by a strong hump in the spectra of the object, which grows larger toward red wavelengths.  
We conclude that this period corresponds to the spin period of the WD and the corresponding frequency peak is denoted as f$_s$.  The one-day alias is very strong in I band, but is easily removed after deconvolution with the spectral window. Other aliases created by uneven distribution of data are also suppressed.  Most of the remaining peaks in the clean power spectrum can be identified with either the spectroscopic (orbital period) marked as f$_o$, or sidebands formed by these two frequencies. Particularly strong are $2\times(f_s - f_o)$; $(f_s - f_o)$ and $2*f_s - f_o$. 
The power spectra corresponding to V and WL light curves are similar regarding the spin period, but the orbital period is not remarkable. In the V period spectrum  there is a strong  $(f_s - f_o)/2$ sideband frequency.

  \begin{figure}[t]
   \setlength{\unitlength}{1mm}
\resizebox{11cm}{!}{
\begin{picture}(100,75)(0,0)
\put   (2,0)   {\includegraphics[width=3.1in,bb=20 160 580 700,clip]{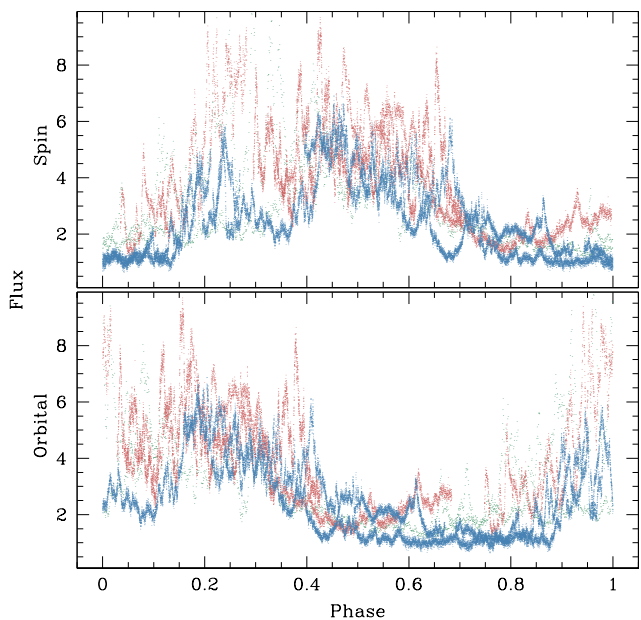}}
  \end{picture}}       
      \caption{Fast photometry. The data from three different nights (two days apart from each other) are folded with orbital (83.6\,m) and spin (81.3\,m) periods.  Different colors indicate different nights. Strong, short variability is probably a result of magnetic "blob" accretion; no signs of WD eclipse can be found.  }
         \label{fig07}
   \end{figure}

Apart from the periodic variability caused by the spin of the magnetic WD and the orbital motion,  there is a huge, erratic variability that is best demonstrated by the high time resolution photometry (Figure\,\ref{fig07})  with fast flares superimposed on a smoother, longer variability. The fast photometric light curves presented in  Figure\,\ref{fig07}  are in fluxes to demonstrate the scale of rapid variability; meanwhile light curves in the remaining figures throughout the paper are in magnitudes (i.e., logarithmic scale).   
While these data provide sufficient time resolution to explore the features of rapid variability, there is no adequate phase coverage to determine the orbital or spin period accurately. 
 
The high-speed light curve folded with the orbital period in the bottom panel of Figure\,\ref{fig07} shows a visually better recurrence of fast-paced features   than that folded with the spin period and presented in the top panel. However a dip just prior to phase 0.4 in the spin-period folded light curve may indicate a self eclipse of the weaker accreting pole.  We could not identify any repetitive luminosity drop corresponding to the presence of an eclipse of stellar components  in the light curves.  The absence of the eclipse constrains the inclination angle of the system to $i < 72\deg$, but provides little information otherwise. 
No periodic signal is detected at higher frequencies,  indicating that the fast and sporadic variability is  probably due to the erratic nature of the accretion flow. 
   
\begin{figure} \setlength{\unitlength}{1mm}
\resizebox{11cm}{!}{
\begin{picture}(100,75)(0,0)
\put   (2,0){\includegraphics[width=3.3in,bb=0 140 620 720,clip]{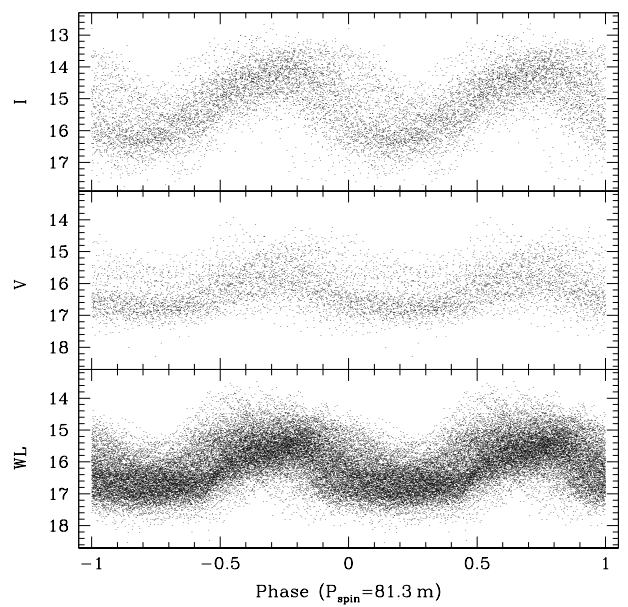}}
  \end{picture}}    
      \caption{Light curves of \igr\ composed of all available data in the white WL, V,  and I bands from bottom panel to top. }
         \label{fig08}
   \end{figure}

The light curves folded with the P$_{s}=81.29$m period are presented in Figure\,\ref{fig08}.  The best defined light curve is a  tide-like structure in the I band presented in the top panel.  Two remarkable features of the plot are a large scatter of the points and a non-sinusoidal form of the curves. The former is not surprising since there is a huge amplitude, rapid variability   
around the brightness maxima revealed by  the  fast photometry. This is also partially due to the brightness variability on a longer timescale. The presence of two periods modulates the light curve with the beat period.   Examples are provided in the top panels of Figure\,\ref{fig05}, where the longer trend is very notable in an unfolded light curve of individual nights.
That allows us not only to fit the  spin period,   but also to fit the trace modulation  with a  longer  period corresponding to the  $f_{2(f_s-f_o)} = 0.976$ frequency. 
The amplitude of the spin period is twice as large as the longer trend. Our discussion of Figures\,\ref{fig08} and \ref{fig05} in conjunction with the RVs and orbital modulation continues in  section\,\ref{sec:osp}.

A signal was detected in  X-rays with a $82.7\pm1.35$\,m  period. This period 
is in the middle and within the errors of two periods detected in the optical domain. The 81.3\m  \ period, however, seems a more natural periodicity to be observed from the magnetic WD beaming collimated X-rays as it spins.  The origin of the longer 101.4\,m period, which was also found in the X-ray data \citep{2013MNRAS.435.2822B}, is not clear. It might be a sideband of the orbital period, but then other sidebands are expected to be seen, too. The X-ray observations do not provide sufficient coverage to sort out these differences or claim reality of other periods.
Interestingly, the 101\,m period also shows up in a least-squares periodogram \citep{1976Ap&SS..39..447L}  of fast photometry (not presented here). However the DFT power spectrum shows a symmetric forest of strong lines with two basic frequencies $17.2\pm n$ and $17.7\pm n$ , where $n=1,2,3, etc$.  No significant signal appears around 101\,m. 

\subsection{Footprint of diverging  orbital and spin periods in the data  }
\label{sec:osp}

\begin{figure}[t]
\resizebox{11cm}{!}{
\begin{picture}(100,75)(0,0)
\put   (0,0)  {\includegraphics[width=1.1in,bb=20 150 600 700,clip]{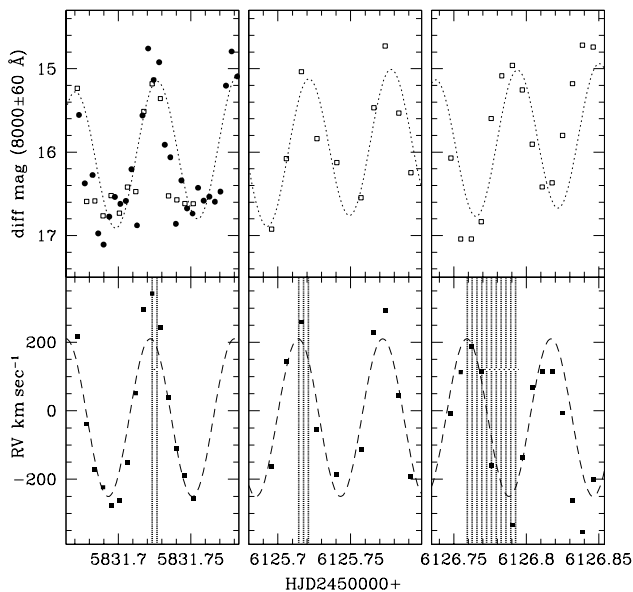}}            
 \end{picture}}     
      \caption{ Bottom panels: Measurements of the RV  of \igr\  at three separate epochs  (filled squares) with a sine fit corresponding to the 
      $f_o$  orbital frequency.
Top panels:   Measurements of  the continuum flux in a band around 8000\,\AA\ (open squares) and I-band magnitudes (filled dots) at the same epochs.  The data are fitted with a sum of sine functions with  $f_s+{2(f_s-f_o)}$ frequencies corresponding to the   
spin  and strongest beat  
frequency. 
The shaded strips in the bottom panels denote the varying shift between phases (maximum RV vs. maximum brightness) caused by the difference of the orbital and spin frequencies.   }
         \label{fig05}
   \end{figure}

Figure\,\ref{fig05}  illustrates how the continuum hump appears at the different orbital phases due to the asynchronism. In the bottom three panels of Figure\,\ref{fig05} measurements of the RVs are presented  against the photometric magnitudes (at the top panels) in three different epochs. The magnitudes are obtained from  photometry when available (filled dots in the top left panel) or from the spectrophotometric fluxes converted to differential magnitudes (open squares).  To obtain the latter, the spectral fluxes were integrated  in a relatively narrow $60\,\AA$ intervals  centered  at  the $\lambda\, 8000\,\AA$ roughly corresponding to the I band,  where the variability of the continuum is the strongest.   Photometric and spectrophotometric measurements obtained at the same time are presented together in the top left panel Figure\,\ref{fig05}.  The errors of spectrophotometric measurements are difficult to assess since they depend on long-slit spectral calibration. We consider that the spectrophotometric data are in reasonable accordance with  the precise photometric data.   The standard deviations of both sets of data  from the $sine$-fit are similar and are much less that the overall variability of the object.  

Apparently, there is a trend in the data besides the obvious large amplitude variability, which we identified with the spin period. Hence, the  curve fitted to the data is a sum of two $sine$ functions: one with the spin frequency  $f_s=17.715$ cycles/day and 0.85\,mag amplitude and the other  with the  ${2(f_s-f_o)}=0.976$ cycles/day corresponding to the strongest beat frequency (see Figure\,\ref{fig06}). 
The beat frequency has a smaller 0.44\,mag amplitude. 
There is a clear displacement of the light and RV curves in the top and bottom panels from epoch to epoch. The shaded strips in each panel denote the time difference of the maximum RV occurrence to the moments of maximum brightness.  The sine curve in the bottom panels has a period determined from the RV fitting and is the same as in  Figure\,\ref{fig03}. The sine curve in the top panels has a shorter period, corresponding to the spin period. 

This displacement means that the  observer has a constantly changing view on the magnetic pole and the magnetically controlled part of the accretion stream. 

\section{Discussion}
\label{discus}

\subsection{Asynchronous polar interpretation}
\label{discussion}

There is no doubt that the object of the study is a magnetic CV. Its periods, optical and  X-ray spectral, and photometric behavior are good enough evidence for that. The spectral shape of the optical bright phase could be well interpreted as a cyclotron continuum (see the evaluation in the next section). 

\begin{figure}[t]\resizebox{11cm}{!}{
\begin{picture}(100,70)(0,0)
\put   (-3,70) {\includegraphics[width=2.35cm,angle=-90,bb = 20 30 540 720,clip=]{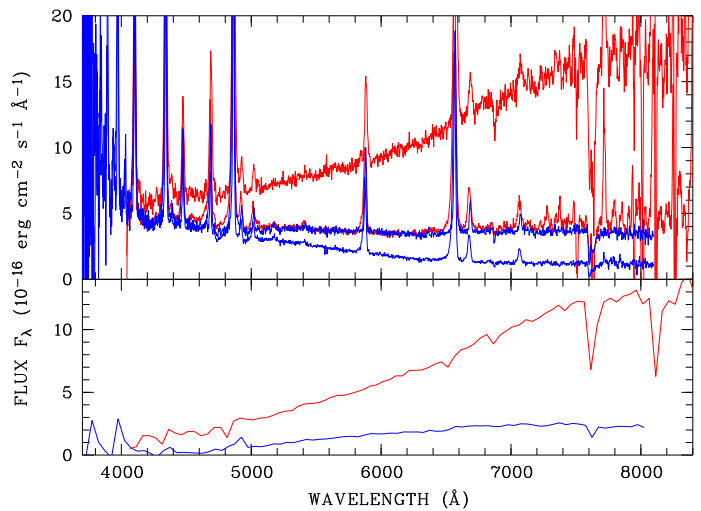}}
 \end{picture}}  
   \caption{Top  panel: Low-resolution spectra of \igr\ obtained in 2011 (blue) and 2012
     (red) at maximum and minimum brightness. Bottom panel: Cyclotron spectra 
in 2011 and 2012 are shown.}
   \label{fig:cycspec} 
\end{figure}

However the system is not an ordinary polar.  
The photometric period that we identify with the spin period of the magnetic WD  is 2.8\%  shorter than the spectroscopic period; we think the latter  reflects the orbital period of the system  ($\frac{P_{o} - P_{s}}{ P_{o}} =0.028$). This is one of the extreme cases of asynchronism \citep{2016MNRAS.458.1833P}.  
As a result of asynchronism the observer looks  at the magnetic pole(s) under constantly changing  angle  with respect to  the orbital phase. Also the coupling region bound to the binary frame changes  its position regarding the magnetic pole (or dipoles) of the asynchronously rotating WD.   Hence the intensity of accretion changes with the position angle, brightness of the system varies, and the cyclotron spectrum varies as perceived by the observer. It is demonstrated convincingly in Figure\,\ref{fig05}. 
Modeling the cyclotron spectrum is a difficult task because of its ever-changing pattern. 

\subsection{Cyclotron spectroscopy}

The pronounced brightness variability on the spin period of the WD is
naturally  explained in terms of cyclotron radiation from one accreting 
pole. The best available representative spectra  at brightness maximum and minimum for the nights September 9, 2011 and July 7, 2016
were chosen to study the cyclotron contribution. These spectra are shown  in
Fig.~\ref{fig:cycspec} with blue and red colors, repectively. In 2012 the overall brightness of the source was
considerably higher so that the minimum spectrum in 2012 is almost a carbon
copy of the maximum spectrum from 2011. The steep
rise toward long wavelength at spin-phase maximum is common to both occasions. 

The difference spectrum
between maximum and minimum is regarded as cyclotron spectrum and shown for
the two occasions in the bottom panel of the same figure. Before subtraction
the strong emissions lines were fitted with Gaussians and removed. After
subtraction the resulting spectra were rebinned to 50\,\AA\ to remove
the high-frequency noise as well. Some residuals due to imperfect subtraction of the
asymmetric emission lines (in particular at wavelengths below 4400\,\AA) and
the non-availability of a telluric absorption spectrum (atmospheric A-band at
7600\,\AA) are left in the spectra. Apart from those, the cyclotron spectra
display a smooth increase toward long wavelengths. It seems as if the maximum
spectral flux occurs around 8000\,\AA\ in 2011 but the spectral maximum might
not be covered by the observations. In 2012 the spectral maximum seems to
occur at wavelengths longer than 8000\,\AA.

There are no individual spectral features that could be unequivocally
associated with the magnetic field in the accretion region: neither a halo
Zeeman  absorption line nor individual cyclotron harmonics. This fact,
together with the red cyclotron spectrum, points toward a relatively low field
strength. The spectral range covered by our observations then corresponds to
the high-harmonic range in which individual cyclotron harmonics overlap strongly
and form a quasi-continuum. If one assumes that the spectral maximum, which
indicates the change from an optically thick Rayleigh Jeans 
to an optically thin
cyclotron spectrum, occurs at 8000\,\AA and that this turnover corresponds to
the eighth cyclotron harmonic, the implied field strength would be about 
16\,MG.  The true field strength could be 35\%\ lower, as argued by  \citet{1997A&A...326..195S},
for the cyclotron spectrum of the AP  CD\,Ind, which is very similar
to that of \igr. In the abovementioned paper the cyclotron spectra of other low 
field polars, such as BL\,Hyi, EP\,Dra, and V393\,Pav, and their similarity to that of CD\,Ind are presented and discussed. 
Further examples are EF\,Eri and V2301\,Oph \citep{1995MNRAS.273...17F,1996MNRAS.282..218F}. 

It appears very unlikely that the
field is larger than about 20\,MG. High accretion rate polars at those field 
strengths typically display a much bluer cyclotron spectrum \citep[cf. MR Ser, V834
Cen][]{1993A&A...278..487S,1990A&A...238..173S} 
The cyclotron lines are probably shifted further down to the infrared as substantiated by large infrared excess 
\citep[Figure 10 of][]{2013MNRAS.435.2822B}.

\subsection{Emission lines composition}

 The high energy  beam from the magnetic pole ionizes the gas in the magnetically controlled stream, which emits the bulk of emission. 
A small fraction of the H$_\alpha$\ line also originates  from the irradiated face of the secondary star indicating a low temperature of the irradiation. 
This was demonstrated vividly by the new, inside-out tomograms (Figure\,\ref{fig:05} top right panels). 
It is worth mentioning that in several polars the irradiated secondary even produces He\,{\sc ii}.  In this case the irradiation of the secondary is 
mild in terms of both contribution and intensity and is observed primarily in H$_\alpha$.
 
The phasing of emission lines is appropriate 
to the proposed interpretation described in detail by \citet{1999MNRAS.304..145H}.   Particularly, the narrow component of the H$_\alpha$\ 
line corresponds to the top panel of \citet[Figure 5 therein]{1999MNRAS.304..145H}, while the broad component, which they call 
the accretion curtain, corresponds to their bottom panel in the figure. In the inside-out projection the crescent-shaped spot at the bottom of corresponding maps reflects  
the presence of that curtain, since it concurs within the area where the  magnetic trajectories (black dotted lines) intercept the matter from the ballistic trajectory (red dotted lines).

The horizontal stream, or the ballistic part of the stream, is practically not visible in H$_\alpha$, 
but becomes visible in H$_\beta$\ inside-out Doppler map.
It is common in polars to see either all or only some components, depending on the location of the magnetic poles and the orientation of the beam. Depending on the level of ionization 
it also can be seen in various species of emission lines.  

Rapid photometry 
shows that the accretion is not smooth but is inhomogeneous and clumpy, which is today a well-established concept  for polars that was proposed by  \citet{1982A&A...114L...4K}. 
In the case of \igr\ this unsteadiness of accretion flow is exaggerated  by the asynchronism. However, the blobby accretion model
was proposed to explain the "softness" of X-ray radiation of polars \citep[][and references therein]{2014EPJWC..6403001W} even though \igr\ is rather hard source.  The soft component is possibly shifted into the unobservable UV range and/or is absorbed  within the systems. 
Actually, IPs are supposed to be harder emitters (in the X-ray), but currently many of these IPs show a soft BB component,  which is a characteristic initially thought to be peculiar of polars only. 
Moreover, observations of polars with XMM-Newton proved that an increasing number of these objects do not show this soft emission. It is not clear why this is the case \citep[][and references therein]{2012A&A...542A..22B}.

\section{Conclusions}
\label{conclusions}

We identified the {\sl INTEGRAL} source IGR\,J19552+0044 as a new asynchronous magnetic CV or polar. Direct evidence of its magnetic nature through either Zeeman or resolved cyclotron lines or by means of (spectro-)polarimetry is outstanding.
Based on optical photometric and spectroscopic observations we determined the orbital and WD spin periods of the object to be 83.6 and 81.3 min, respectively. The 2.8\% rate of asynchronism is among the largest observed in a few similar objects. We only have an estimate of a moderate $\approx 16$\,MG field strength of the WD.  This estimate agrees well with the assessment  of infrared excess by \citet{2013MNRAS.435.2822B}. Doppler tomography of emission lines confirm the small size of the WD magnetosphere, showing the accretion stream treading area all the way down the ballistic trajectory, close to the WD. Very fast photometry demonstrates large variability on  very short timescales, which is consistent with the generally accepted point of view that the matter hits the magnetic pole in the form of blobs, rather than a fluid stream. The source of the spin period deviation from the orbital period in APs is not established yet and IGR\,J19552+0044 sheds little light on that. But the growing number of discovered APs indicate that it is not as rare as originally thought.

\begin{acknowledgements}

GT and SZ acknowledge PAPIIT grants IN-108316/IN-100617 and CONACyT grants 166376; 151858 and CAR 208512 for resources provided toward this research. JT acknowledges the  NSF grant AST-1008217.
      
\end{acknowledgements}


\begin{thebibliography}{}

\bibitem[Bernardini et al.(2012)]{2012A&A...542A..22B} Bernardini, F., de Martino, D., Falanga, M., et al.\ 2012, \aap, 542, A22

\bibitem[Bernardini et al.(2013)]{2013MNRAS.435.2822B} Bernardini, F., de Martino, D., Mukai, K., et al.\ 2013, \mnras, 435, 2822 

\bibitem[Bird et al.(2006)]{2006ApJ...636..765B} Bird, A.~J., Barlow, E.~J., Bassani, L., et al.\ 2006, \apj, 636, 765 

\bibitem[Boyd et al.(2014)]{2014SASS...33..163B} Boyd, D., Patterson, J., Allen, W., et al.\ 2014, Society for Astronomical Sciences Annual Symposium, 33, 163 

\bibitem[Campbell \& Schwope(1999)]{1999A&A...343..132C} Campbell, C.~G., \& Schwope, A.~D.\ 1999, \aap, 343, 132 


\bibitem[Campbell(1985)]{1985MNRAS.215..509C} Campbell, C.~G.\ 1985, 
\mnras, 215, 509 

\bibitem[Cropper(1990)]{1990SSRv...54..195C} Cropper, M.\ 1990, \ssr, 54, 195 

\bibitem[de Miguel et al.(2016)]{2016MNRAS.457.1447D} de Miguel, E., Patterson, J., Cejudo, D., et al.\ 2016, \mnras, 457, 1447

\bibitem[Ferrario et al.(1995)]{1995MNRAS.273...17F} Ferrario, L., Wickramasinghe, D., Bailey, J., \& Buckley, D.\ 1995, \mnras, 273, 17 

\bibitem[Ferrario et al.(1996)]{1996MNRAS.282..218F} Ferrario, L., Bailey, J., \& Wickramasinghe, D.\ 1996, \mnras, 282, 218 

\bibitem[Ferrario \& Wehrse(1999)]{1999MNRAS.310..189F} Ferrario, L., \& Wehrse, R.\ 1999, \mnras, 310, 189 

\bibitem[Ferrario et al.(2015)]{2015SSRv..191..111F} Ferrario, L., de Martino, D., \& G{\"a}nsicke, B.~T.\ 2015, \ssr, 191, 111

\bibitem[Harrison \& Campbell(2016)]{2016MNRAS.tmp..737H} Harrison, T.~E., \& Campbell, R.~K.\ 2016, \mnras,  

\bibitem[Heerlein et al.(1999)]{1999MNRAS.304..145H} Heerlein, C., Horne, K., \& Schwope, A.~D.\ 1999, \mnras, 304, 145 

\bibitem[King 
\& Whitehurst(1991)]{1991MNRAS.250..152K} King, A.~R., \& Whitehurst, R.\ 1991, \mnras, 250, 152

\bibitem[Knigge(2006)]{2006MNRAS.373..484K} Knigge, C.\ 2006, \mnras, 373, 484  

\bibitem[Kotze et al.(2016)]{2016A&A...595A..47K} Kotze, E.~J., Potter, S.~B., \& McBride, V.~A.\ 2016, \aap, 595, A47 

\bibitem[Kuijpers \& Pringle(1982)]{1982A&A...114L...4K} Kuijpers, J., \& Pringle, J.~E.\ 1982, \aap, 114, L4 

\bibitem[Lenz \& Breger(2005)]{2005CoAst.146...53L} Lenz, P., \& Breger, M.\ 2005, Communications in Asteroseismology, 146, 53

\bibitem[Lomb(1976)]{1976Ap&SS..39..447L} Lomb, N.~R.\ 1976, \apss, 39, 447  

\bibitem[Marsh \& Horne(1988)]{1988MNRAS.235..269M} Marsh, T.~R., \& Horne, K.\ 1988, \mnras, 235, 269 

\bibitem[Marsh \& Schwope(2016)]{2016ASSL..439..195M} Marsh, T.~R., \& Schwope, A.~D.\ 2016, Astronomy at High Angular Resolution, 439, 195

 \bibitem[Masetti et 
al.(2010)]{2010A&A...519A..96M} Masetti, N., Parisi, P., Palazzi, E., et al.\ 2010, \aap, 519, A96

\bibitem[Norton et al.(2004)]{2004ApJ...614..349N} Norton, A.~J., Wynn, 
G.~A., \& Somerscales, R.~V.\ 2004, \apj, 614, 349 

\bibitem[Pagnotta \& Zurek(2016)]{2016MNRAS.458.1833P} Pagnotta, A., \& Zurek, D.\ 2016, \mnras, 458, 1833 

\bibitem[Rea et al.(2016)]{2016arXiv161104194R} Rea, N., Coti Zelati, F., Esposito, P., et al.\ 2016, arXiv:1611.04194

\bibitem[Reichart et al.(2005)]{2005NCimC..28..767R} Reichart, D., Nysewander, M., Moran, J., et al.\ 2005, Nuovo Cimento C Geophysics Space Physics C, 28, 767 


\bibitem[Roberts et al.(1987)]{1987AJ.....93..968R} Roberts, D.~H., Lehar, J., \& Dreher, J.~W.\ 1987, \aj, 93, 968

\bibitem[Skillman \& Patterson(1993)]{1993ApJ...417..298S} Skillman, D.~R., \& Patterson, J.\ 1993, \apj, 417, 298 

\bibitem[Schwarz et al.(1998)]{1998A&A...338..465S} Schwarz, R., Schwope, A.~D., Beuermann, K., et al.\ 1998, \aap, 338, 465

\bibitem[Schwarz et al.(2004)]{2004ASPC..315..230S} Schwarz, R., Schwope, A.~D., Staude, A., et al.\ 2004, IAU Colloq.~190: Magnetic Cataclysmic Variables, 315, 230

\bibitem[Schwarz et al.(2007)]{2007A&A...473..511S} Schwarz, R., Schwope, A.~D., Staude, A., et al.\ 2007, \aap, 473, 511 

\bibitem[Schwope \& Beuermann(1990)]{1990A&A...238..173S} Schwope, A.~D., \& Beuermann, K.\ 1990, \aap, 238, 173 

\bibitem[Schwope et al.(1993)]{1993A&A...278..487S} Schwope, A.~D., Beuermann, K., Jordan, S., \& Thomas, H.-C.\ 1993, \aap, 278, 487 

\bibitem[Schwope et al.(1997)]{1997A&A...319..894S} Schwope, A.~D., Mantel, K.-H., \& Horne, K.\ 1997, \aap, 319, 894

\bibitem[Schwope et al.(1997)]{1997A&A...326..195S} Schwope, A.~D., Buckley, D.~A.~H., O'Donoghue, D., et al.\ 1997, \aap, 326, 195 

\bibitem[Spruit(1998)]{1998astro.ph..6141S} Spruit, H.~C.\ 1998, arXiv:astro-ph/9806141

\bibitem[Stockman et al.(1988)]{1988ApJ...332..282S} Stockman, H.~S., Schmidt, G.~D., \& Lamb, D.~Q.\ 1988, \apj, 332, 282 

\bibitem[Thorstensen \& Halpern(2013)]{thorhalpern13} Thorstensen, J.~R., \& Halpern, J.\ 2013, \aj, 146, 107

\bibitem[Warner(1995)]{1995ASPC...85....3W} Warner, B.\ 1995, Magnetic Cataclysmic Variables, 85, 3

\bibitem[Wickramasinghe(2014)]{2014EPJWC..6403001W} Wickramasinghe, D.\ 2014, European Physical Journal Web of Conferences, 64, 03001

\end{thebibliography}
\end{document}